\documentclass[useAMS,usenatbib]{mn2e}
\usepackage{graphicx} 
\usepackage{color}
\usepackage{float}
\usepackage{caption}
\usepackage{placeins}
\usepackage{amsmath}
\usepackage{caption}
\usepackage{multicol}

\def\be{\begin{equation}} 
\def\ee{\end{equation}}

\def\mpc{\,{\rm {Mpc}}} 
 
\def\kms{\,{\rm {km\, s^{-1}}}}

\def\gsim{\lower.5ex\hbox{\gtsima}} 
\def\lsim{\lower.5ex\hbox{\ltsima}} \def\gtsima{$\; \buildrel > \over \sim \;$} \def\ltsima{$\; \buildrel < \over \sim \;$} \def\prosima{$\; 
\buildrel \propto \over \sim \;$} \def\gsim{\lower.5ex\hbox{\gtsima}} 
\def\lsim{\lower.5ex\hbox{\ltsima}} 
\def\simgt{\lower.5ex\hbox{\gtsima}} 
\def\simlt{\lower.5ex\hbox{\ltsima}} 
\def\simpr{\lower.5ex\hbox{\prosima}} \def\la{\lsim} \def\ga{\gsim} 
  
 \def\gtsima{$\; \buildrel > \over \sim \;$} 
\def\ltsima{$\; \buildrel < \over \sim \;$} 
\def\gsim{\lower.5ex\hbox{\gtsima}} 
\def\lsim{\lower.5ex\hbox{\ltsima}} 
\def\simgt{\lower.5ex\hbox{\gtsima}} 
\def\simlt{\lower.5ex\hbox{\ltsima}} 
\def\simpr{\lower.5ex\hbox{\prosima}} 
\def\la{\lsim} 
\def\ga{\gsim}

\def\E3{{\cal E}_{\rm g}^{III}}

\def\Msun{\rm M_\odot}
\def\Zsun{\rm Z_\odot}

\def\Msun{\rm M_\odot}
\def\Zsun{\rm Z_\odot}
\def\M*{M_*}
\def\Z*{Z_*}
\def\L*{L_*}

\def\kev{\,\rm{keV}}

\def\mpc{\rm Mpc}
\def\kms{\rm km\, s^{-1}}
\def\c4{\rm CIV}
\def \muv{\rm M_{UV}}

\title[WDM constraints using IGM metal enrichment]{Probing the nature of Dark Matter through the metal enrichment of the intergalactic medium} 
\author[Bremer et. al.]{Jonas Bremer$^{1}$\thanks{bremer@astro.rug.nl}, Pratika Dayal$^1$ \& Emma V. Ryan-Weber$^2$\\ 
$^{{1}}$ Kapteyn Astronomical Institute, University of Groningen, PO Box 800, 9700 AV Groningen, The Netherlands\\
$^{2}$ Centre for Astrophysics \& Supercomputing, Swinburne University of Technology, PO Box 218, Hawthorn, VIC 3122, Australia \\
}

\begin{document} 
 
\date{} 

\maketitle

\begin{abstract}
We focus on exploring the metal enrichment of the intergalactic medium (IGM) in Cold and Warm (1.5 and 3 keV) Dark Matter (DM) cosmologies, and the constraints this yields on the DM particle mass, using a semi-analytic model, {\it Delphi}, that jointly tracks the Dark Matter and baryonic assembly of galaxies at $z \simeq 4-20$ including both Supernova and (a range of) reionization feedback (models). We find that while $\muv \gsim -15$ galaxies contribute half of all IGM metals in the Cold Dark Matter model by $z \simeq 4.5$, given the suppression of low-mass halos, larger halos with $\muv \lsim -15$ provide about 80\% of the IGM metal budget in 1.5 keV Warm Dark Matter models using two different models for the metallicity of the interstellar medium. Our results also show that the only models compatible with two different high-redshift data sets, provided by the evolving Ultra-Violet luminosity function at $z \simeq 6-10$ and IGM metal density \citep[e.g.][]{simcoe2011}, are standard Cold Dark Matter and 3 keV Warm DM that do not include any reionization feedback; a combination of the UV LF and the \citet{diaz2016} points provides a weaker constraint, allowing Cold and 3 keV and 1.5 keV Warm DM models with SN feedback only, as well as CDM with complete gas suppression of all halos with $v_{circ} \lsim 30\, \kms$. Tightening the error bars on the IGM metal enrichment, future observations, at $z \gsim 5.5$, could therefore represent an alternative way of shedding light on the nature of Dark Matter.
\end{abstract}


\section{Introduction}
The particle nature of Dark Matter (DM) remains one of the key outstanding problems in the field of physical cosmology.  The standard Lambda Cold Dark Matter ($\Lambda$CDM) cosmological model has now been successfully tested using the large scale ($10-100$ Mpc) structure of the Universe inferred from the Cosmic Microwave Background (CMB), the Lyman Alpha forest, galaxy clustering and weak lensing  \citep[see e.g.][]{weinberg2015}. However, the elegance of this picture is marred by the fact that CDM seems to exhibit an excess of power on small-scales \citep[summarised in e.g.][]{weinberg2015, popolo2017}. This ``small-scale crisis'' manifests itself in the observed lack of theoretically predicted satellites of the Milky Way \citep[``the missing satellite problem'';] []{moore1999, klypin1999},  DM halos being too dense as compared to observations \citep[``the core-cusp problem'';][]{moore1999b, navarro1997} and in the lack of  theoretically predicted massive satellites of the Milky Way \citep[``too big to fail problem'';][]{BoylanKolchin11,BoylanKolchin12}. Although some of these problems can be solved purely through the effects of baryonic feedback including, but not limited to, the effects of Supernovae (SN) and parent-satellite interactions \citep{Koposov, popoloparsatint,garrisonkimmel,madaudmheating,pearrubia, maccio2012, dicintio2014, governato2012, governato2015, Silk}, an alternative route focuses on questioning the cold nature of Dark Matter itself.  One such alternative candidate is provided by Warm Dark Matter (WDM) with particle masses $m_x \sim \mathcal{O}$(keV) \citep[e.g.][]{Bode2001}. In addition to its particle-physics motivated nature, the WDM model has been lent support by the observations of a 3.5 keV line from the Perseus cluster that might arise from the annihilation of light sterile neutrinos into photons \citep{bulbul2014,boyarsky2014,cappelluti2017}. However, other works \citep{maccio,schneider} caution that the power-suppression arising from WDM makes it incompatible with observations, leaving the field open to other models including fuzzy CDM consisting of ultra light $\mathcal{O}$(10$^{-22}$eV) boson or scalar particles \citep{fuzzy2000,fuzzydetail2017,fuzzysubstr}, self-interacting DM \citep{Spergel,rocha2013,vogelsberger2014} and decaying DM \citep{wang2014}. The most recent estimates of the (thermally decoupled) WDM particle mass range between $m_x \gsim 2-2.9 \kev$ \citep[using Milky Way dwarf satellites;][]{Kennedy,Jethwa}, $m_x \gsim 2.9-5.3\kev$ \citep[from Lyman Alpha forest statistics;][]{Viel,Baur,Irsic}, $m_x \gsim 1.3-3\kev$ \citep[from reionization;][]{Tan, Lopez}, $m_x>1.8 \kev$ \citep[from ultra-deep ultraviolet luminosity functions at $z \simeq 2$;][]{Menci2}, $m_x\gsim 1.6 \kev$ \citep[from high-$z$ Gamma Ray Bursts;][]{Souza} and  $m_x \gsim 1-2.1 \kev$ \citep[by modelling high-$z$ galaxies and gravitational lenses;][]{Pacucci,Kaiki,Menci,Birrer}. A number of works have also shown how forthcoming observations with, for example, the James Webb Space Telescope ({\it JWST}), can be used to differentiate between $m_x\lsim 1.5 \kev$ and $m_x\gsim 3\kev$ WDM using the redshift-dependent growth of the  stellar mass density \citep{dayal2014_wdm1}, stellar mass-halo mass relations \citep{dayal2017} and high-$z$ Direct Collapse Black Holes \citep{dayal2017b}.

In this {\it proof-of-concept} work our aim is to, firstly, study the metal-enrichment of the intergalactic medium (IGM) at high-$z$ ($z \gsim 4$) in both cold and warm matter cosmologies and, secondly, check if the IGM metal enrichment can be used to place constraints on the WDM particle mass.  Our motivation arises from the fact that, with their shallow potentials, galaxies with low halo masses ($\lsim 10^{9.5}\Msun$) are expected to be the dominant contributors to the IGM metal budget at high-$z$ \citep[e.g.][]{oppenheimer2009, shen2013, finlator2015, diaz2015, garcia2017}. Therefore, the increasing lack of such low-mass halos, due to an increasing suppression of small-scale power, with decreasing $m_x$ will lead to both a delay and a decrease in the IGM metal-enrichment at early cosmic epochs.  

\begin{figure}
\centering{\includegraphics[width = 0.45\textwidth]{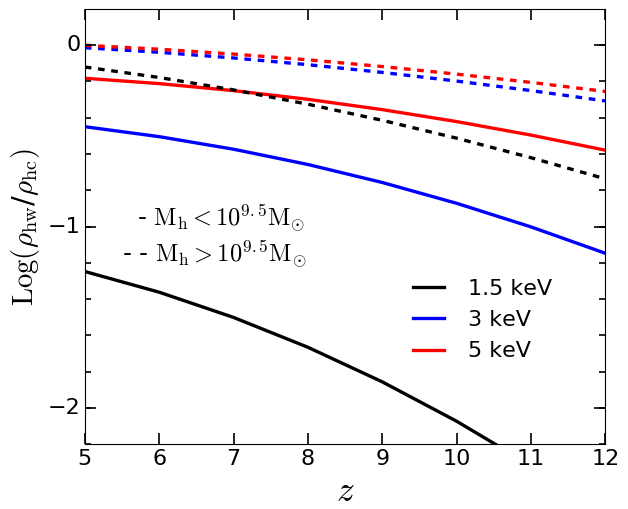}}
\caption{As a function of redshift, we show the (Log) cumulative mass density bound in WDM halos ($\rho_{\rm hw}$) relative to CDM ($\rho_{\rm hc}$) for three different WDM masses: $1.5\kev$ (black lines), $3\kev$ (blue lines) and $5\kev$ (red lines) respectively.  Solid and dashed lines show the mass bound in halos with $M_h \lsim 10^{9.5}\Msun$ and $M_h \gsim 10^{9.5}\Msun$, respectively.}
\label{missingdm} 
\end{figure}

We illustrate this point using Fig. \ref{missingdm} that shows the cumulative mass density contained in bound DM halos in three different WDM models, with $m_x=1.5,3$ and $5 \kev$, with respect to CDM. Firstly, focusing at $M_h \gsim 10^{9.5}\Msun$ halos we see that the $5\kev$ WDM particle is heavy enough to have assembled 55\% of the total mass density of CDM halos by $z \simeq 12$, increasing  to $\sim 100\%$ by $z \simeq 5$. Given its low mass, and correspondingly large suppression of power on small scales, the 1.5 keV WDM model has only assembled about $18\%$ of the halo mass density compared to CDM by $z\simeq 12$, increasing to $\sim 76\%$ by $z \simeq 5$; as expected, the $3 \kev$ model straddles the range between these two extremes, lying close to the $5\kev$ WDM results. On the other hand, there is significant bound DM mass missing when considering low-mass halos with $M_h \lsim 10^{9.5}\Msun$: indeed, the $1.5 \kev$ WDM model assembles $<1\%$ of the total CDM mass in such halos at $z \simeq 12$, rising only to $\sim 6\%$ by $z \simeq 5$. This dearth of bound halos naturally implies a dearth in metal-production and, by extension,  the metal-enrichment of the IGM.  As expected, the  bound mass fraction increases with $m_x$ to $\sim 26\%$ at $z \simeq 12$ and is as high as $66\%$ at $z \simeq 5$ for 5 keV WDM. 

We start by describing the theoretical model in Sec. \ref{sec_model}. We quantify the impact of both SN feedback and (a suite of) reionization feedback scenarios on, both, the stellar/gas content of early galaxies in Sec. \ref{sec_fb} before evaluating the metal enrichment of the IGM and comparing to the observed IGM metallicities in Sec.\ref{sec_IGMenrichment}. 
Throughout this paper, we use the latest cosmological parameters  as measured by the {\it Planck} satellite \citep{planck2015} such that ($\Omega_{\rm m },\Omega_{\Lambda}, \Omega_{\rm b}, h, n_s, \sigma_8) = (0.3089,0.6911,0.0486, 0.6774, 0.9667, 0.8159)$ and quote all quantities in comoving units unless stated otherwise. Here, $\Omega_{\rm m },\Omega_{\Lambda}, \Omega_{\rm b}$ represent the density parameters for matter, Dark Energy and baryons, respectively, $h$ is the Hubble value, $n_s$ is the spectral index of the initial density perturbations and $\sigma_8$ represents the root mean square density fluctuations on scales of $8 h^{-1}$ cMpc.
 
\section{The Theoretical model}
\label{sec_model}
The calculations presented in this work are based on the semi-analytic model {\it Delphi} \citep[{\bf D}ark Matter and the {\bf e}mergence of ga{\bf l}axies in the e{\bf p}oc{\bf h} of re{\bf i}onization;][]{dayal2014, dayal2014_wdm1, dayal2017, dayal2017b} that jointly tracks the DM and baryonic assembly of  high-$z$ ($z \sim 4-20$) galaxies.  We start by generating modified binary merger-trees with accretion \citep{parkinson2008} for 800 (4000) galaxies at $z =4$ in CDM ($1.5\kev$ WDM), uniformly distributed in the halo mass range ${\rm Log}(M_h/\Msun)=9-13$. We use the modifications required to generate merger-trees for WDM presented in \citet{benson2013} that include introducing: (a) a $m_x$-dependent cut-off in the initial power spectrum; (b) using a $m_x$-dependent critical over-density of collapse; (c) using a sharp window function in $k$-space; and (d) using numerically calibrated DM infall rates. Matching to the Sheth-Tormen halo mass function (HMF) at $z=4$ yields the (comoving) number density for each halo which is propagated throughout its merger-tree; we have confirmed the resulting HMFs are in agreement with the Sheth-Tormen HMF at all $z\simeq 4.5-20$. 
  
\begin{figure*}
\center{\includegraphics[scale=0.35]{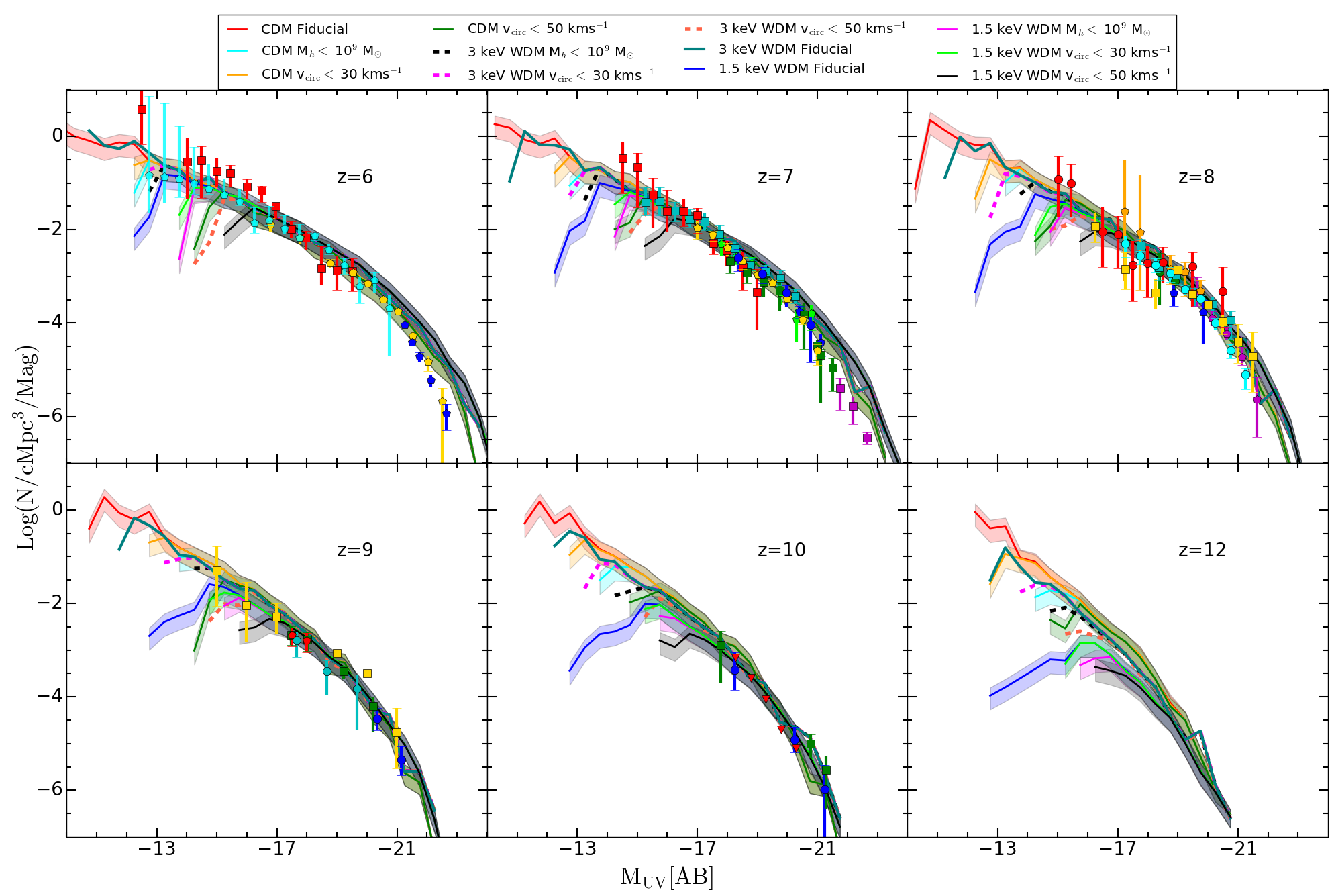}}
\caption{\small The UV Luminosity Functions for CDM,  $3 \kev$ and $1.5 \kev$ WDM for $z\simeq 6-12$, as marked. In each panel, the different lines show results for the four feedback models adopted (see Sec. \ref{sec_model}), as marked in the legend, with the shaded regions showing the 1-$\sigma$ poissonian errors; for clarity, the $3 \kev$ model is shown without errors. In each panel, points show observational data- $z \simeq 6$: \citet[][gold pentagons]{bouwens2015}, \citet[][blue pentagons]{bowler2015}, \citet[][red squares]{livermore2017}, \citet[][cyan pentagons]{bouwens2016b}; $z \simeq 7$: 
\citet[][blue pentagons]{castellano2010}, \citet[][green squares]{mclure2010}, \citet[][blue circles]{oesch2010}, \citet[][green pentagons]{bouwens2011b}, \citet[][gold pentagons]{mclure2013}, \citet[][magenta squares]{bowler2014a}, \citet[][cyan squares]{atek2015}, \citet[][red squares]{livermore2017}; $z \simeq 8$: \citet[][green circles]{bouwens2010a}, \citet[][blue pentagons]{mclure2010}, \citet[][cyan squares]{bouwens2011b}, \citet[][magenta pentagons]{bradley2012}, \citet[][cyan circles]{mclure2013}, \citet[][orange pentagons]{atek2015}, \citet[][red circles]{livermore2017} and \citet[][gold squares]{ishigaki2017}; $z \simeq 9$: \citet[][red pentagons]{mclure2013}, \citet[][cyan hexagons]{oesch2013}, \citet[][green squares]{mcleod2016}, \citet[][blue circles]{bouwens2016}, \citet[][gold squares]{ishigaki2017}; $z \simeq 10$: \citet[][blue circles]{bouwens2015}, \citet[][green squares]{oesch2014} and \citet[][red triangles showing the upper limits]{oesch2014}. }
\label{fig_uvlf} 
\end{figure*}

As for the baryonic physics,  the first  progenitor(s) of any halo are assigned a gas mass that scales with the halo mass through the cosmological ratio such that $M_g = (\Omega_b/\Omega_m) M_h$. A fraction of this gas mass is converted into stars with an effective star formation efficiency ($f_*^{eff}$) that is the minimum between the efficiency that produces enough type II supernova (SNII) energy to eject the rest of the gas, $f_*^{ej}$, and an upper maximum threshold, $f_*$, so that $f_*^{eff} = min[f_*^{ej}, f_*]$.  We calculate the newly formed stellar mass at any $z$ as $M_*(z) = M_g (z) f_*^{eff}$ and the final gas mass at the end of the $z$-step, including that lost in star formation and SN feedback, is then given by $M_{gf}(z) = [M_g(z)-M_*(z)] [1-(f_*^{eff}/f_*^{ej})]$. At each $z$-step we also account for DM that is smoothly accreted from the IGM, making the reasonable assumption that  this is accompanied by  accretion of a cosmological fraction ($\Omega_b/\Omega_m$) of gas mass. 

We use a Salpeter initial mass function \citep[IMF;][]{salpeter1955} between $0.1-100\Msun$ throughout this work. Assuming a fixed metallicity of $0.2 \Zsun$ for all stars, we then use the stellar population synthesis code  {\it Starburst99} \citep{leitherer1999,leitherer2010} to generate the complete spectrum for each galaxy summing over all its entire star formation history. This physical prescription yields model results in excellent agreement with all currently available data-sets for high-$z$ ($z \gsim 5$) galaxies, from the evolving Ultra-violet luminosity function (UV LF) to the stellar mass density (SMD) to mass-to-light ratios to the $z$-evolution of the stellar mass and UV luminosity densities, for both CDM and WDM. We note that the model only uses two mass- and $z$-independent free parameters: to match to observations we require (roughly) 10\% of the SNII energy coupling to the gas ($f_w$) and a maximum (instantaneous) star formation efficiency of $f_* = 3.5\%$. This (SNII feedback only) model is designated as the {\it fiducial model} in what follows.

In this work, we also include the effects of the  Ultra-violet background (UVB) created during reionization which, by heating the ionized IGM to $T \sim 10^4$ K,  can have an impact on the baryonic content of low-mass halos \citep[e.g.][]{okamoto2008, petkova2011, ocvirk2016}. Maintaining the same SNII feedback and $f_*$ parameters as the fiducial model, in this work, we also consider three (maximal) UVB-feedback scenarios in which the gas mass is completely photo-evaporated for halos: (i) below a characteristic halo mass of $M_{h} = 10^{9}\Msun$; (ii) below a circular velocity of $v_{circ} = 30\, \kms$; and (iii) below a circular velocity of  $v_{circ} = 50\, \kms$.  In the latter two cases, the minimum halo mass affected by the UVB increases with decreasing $z$ (since $v_{circ}(z) \propto M_{h}^{0.33} (1+z)^{0.5}$) from $M_h \simeq 10^{8.6}$ to $M_h\simeq 10^{9.1}\Msun$ ($\simeq 10^{9.2}$ to $10^{9.7}\Msun$) from $z \simeq 12$ to 5 for a velocity cut of $v_{circ} = 30\, \kms\, (50 \, \kms)$. Therefore the UV feedback scenario with $M_h = 10^9\Msun$ lies between the constant velocity cut-off cases considered here, lying close to case (iii) at the highest redshifts and slowly tending towards case (ii) by $z \simeq 5$.

Finally, in order to calculate the IGM metal enrichment driven by outflows from these early galaxy populations, we assume gas and metals to be perfectly mixed in the ISM, and carry out calculations for two limiting scenarios: the first, where every galaxy has a fixed metallicity of $Z_{gas}=0.20\Zsun$ and the second where the gas-phase metallicity for each galaxy depends on its stellar mass.


\section{Impact of feedback in Cold and Warm Dark Matter models}
\label{sec_fb}
We now use the model explained above to quantify the impact of internal (SNII) and external (UVB) feedback on galaxy observables, including the evolving UV LF and the SMD, and intrinsic properties, such as the total density of ejected gas mass, for both cold and warm dark matter cosmologies. 

\subsection{Feedback impact on the UV LF}
\label{sec_uvlf}
Quantifying the number density of Lyman Break Galaxies (LBG) as a function of the UV luminosity, the UV LF and its $z$-evolution, offer a robust data-set against which to calibrate the model. As noted above, {\it Delphi} uses two parameters to match to the observed data - an instantaneous star formation efficiency ($f_*=0.035$) and the fraction of SNII energy coupling to gas ($f_w =0.1$) which, broadly, impact the bright and faint ends of the UV LF, respectively. The results of these calculations are shown in Fig. \ref{fig_uvlf}. Starting with CDM, the fiducial model extends to magnitudes as faint as $\muv = -10\, (-12)$ for $z \simeq 5\, (12)$ with a faint-end slope that evolves as $\alpha = -1.75 \, {\rm log} (z) -0.52$ \citep[see also][]{dayal2014}. We note that this model is in excellent agreement with all available observational data at $z \simeq 6-10$; the slight over-prediction of the number density of the rarest brightest $z \simeq 6$ galaxies possibly arises due to our ignoring the effects of dust attenuation for these massive systems. Given that the impact of UV feedback, in suppressing the baryonic content of low-mass halos, progressively increases using a cut-off of $v_{circ} = 30\, \kms$ to $M_h =10^9 \Msun$ to $v_{circ} = 50\, \kms$, we find that the UV LF starts peeling away from the fiducial UV LF at increasing luminosities (decreasing magnitudes) in the same order. Indeed, as seen from Fig. \ref{fig_uvlf}, cutting off at $\muv \sim -12.5$ at $z \simeq 6$, the CDM UV feedback models assuming no gas in halos below $M_h =10^9 \Msun$ and $v_{circ} = 30\, \kms$ are compatible with all available observations except for the faintest $\muv=-12.5$ point at $z \simeq 6$ inferred using lensed Hubble Space telescope (HST) data \citep{livermore2017}. A confirmation of the faint-end slope persistently rising to such faint magnitudes, corresponding to halo masses of about $10^{8.5-9}\Msun$, might be a powerful test of the nature of DM and the impact of feedback on these low-mass systems. However, with its impact on larger halo masses, the $v_{circ} = 50\, \kms$ model naturally cuts-off at higher luminosities corresponding to $\muv \approx -15 \, (-16)$ at $z \simeq 6 \, (12)$ - using current data, we can therefore rule out this maximal UV suppression model. We also find that, although, the halo mass range affected by UV feedback increases by about 0.5 dex between $z \simeq 13-5$, the shift in the UV LF between this range is larger ($\sim 1.5$ magnitudes) than the expected value ($\sim 0.75$) - this is the result of the $L_{UV}/M_*$ value decreasing with decreasing $z$ \citep[see Fig. 7;][]{dayal2014}. Yielding results in accord with CDM down to $\muv \approx -11 \, (-13)$ at $z \simeq 6 \, (12)$, the fiducial 3 keV WDM model is in accord with all available data points; indeed, the 3 keV WDM for complete UV suppression in all halos below $v_{circ} = 30\, \kms$ also matches all available data except the faintest $\muv=-12.5$ point at $z \simeq 6$ \citep{livermore2017}.  

\begin{figure*}
\center{\includegraphics[scale=0.55]{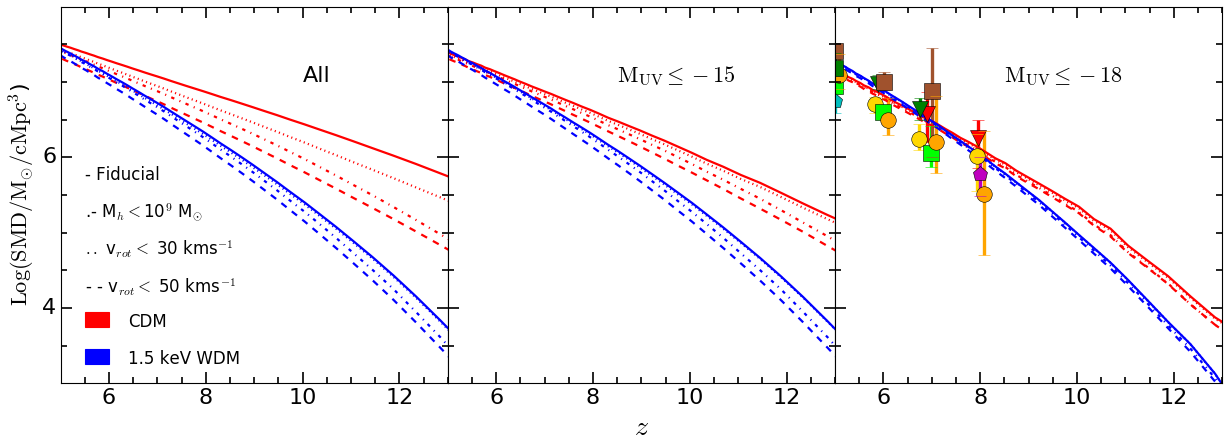}}
\caption{The stellar mass density as a function of redshift for all galaxies (left panel), galaxies with $\muv \lsim-15$ (middle panel) and $\muv \lsim-18$  (right panel). In each panel, the red and blue lines show results for CDM and 1.5 keV WDM, respectively, for the different feedback models noted in the legend. Points show SMD measurements inferred using observations for a limiting magnitude of $\muv\simeq-18$: \citet[][green square]{yabe2009}, \citet[][red triangles]{labbe2010a,labbe2010b}, \citet[][green triangles]{gonzalez2011}, \citet[][cyan pentagon]{lee2012}, \citet[][magenta pentagon]{labbe2013}, \citet[][yellow circles]{stark2013}, \citet[][brown squares]{Duncan2014}, \citet[][light green squares]{Grazian2015} and \citet[][orange circles]{Song2016}.}
\label{fig_smd} 
\end{figure*}

The 1.5 keV fiducial model yields results that are qualitatively the same as the fiducial CDM case down to $\muv \simeq -13$ at $z \simeq 6$ and  
given the increasing lack of low-mass halos with increasing redshift, turns-over at progressively brighter magnitudes with increasing redshift ($\muv \simeq -18$ at $z \simeq 12$). It is interesting to see that the fiducial 1.5 keV model lies close to the CDM $v_{circ} = 50\, \kms$ UV feedback case at $z \simeq 12$, and shifts closer to the CDM $v_{circ} = 30\, \kms$ case by $z \simeq 6$. We also find that, within error bars, the 1.5 keV fiducial model is also in agreement with all available data except for the one $z =6$ data point at $\muv = -12.5$ \citep{livermore2017}. Including the impact of UV feedback, we again find the same trends as CDM, although the magnitude cuts at which the UV LF starts peeling away from the fiducial case correspond to much brighter galaxies. Indeed, unless we modify the baryonic physics for each UV feedback model, we find that current $\muv \gsim-14$ LBG data at $z =6-7$ \citep{livermore2017, bouwens2016b} can effectively be used to rule out ``maximal" UV feedback scenarios. However, we caution that, in principle, only the fraction ($1-Q_{II}$ where $Q_{II}$ is the volume filling factor of ionized hydrogen) of galaxies embedded in ionized regions should be affected by UV feedback at any redshift. This implies that the ``true" (SNII + UV feedback affected) UV LF should lie between the fiducial and ``maximal" UV suppression cases considered here. 

\subsection{Feedback impact on the stellar mass density}
Encoding the total mass locked up in stars, the stellar mass density and its redshift evolution presents a crucial test for any model of galaxy formation. Once that our model free parameters have been fixed by matching to the UV LF as explained above, we study the SMD and compare our theoretical SMD values with observational data. We start by noting that all CDM and 1.5 keV WDM models, both fiducial and including maximal UV feedback, yield SMD results in excellent agreement with observations of $\muv \lsim -18$ galaxies. Although a robust test of our model, this implies that currently observed galaxies can not be used to distinguish between CDM and WDM models, requiring observations to extend down to fainter magnitudes \citep[see also][]{dayal2014}. In what follows, we limit ourselves to studying CDM and 1.5 keV WDM \citep[corresponding to a sterile neutrino mass of $7.6 \kev$;][]{viel2005} given that their comparison should show the largest dearth of halos and hence the largest difference in the SMD.

\begin{figure*}
\center{\includegraphics[scale=0.62]{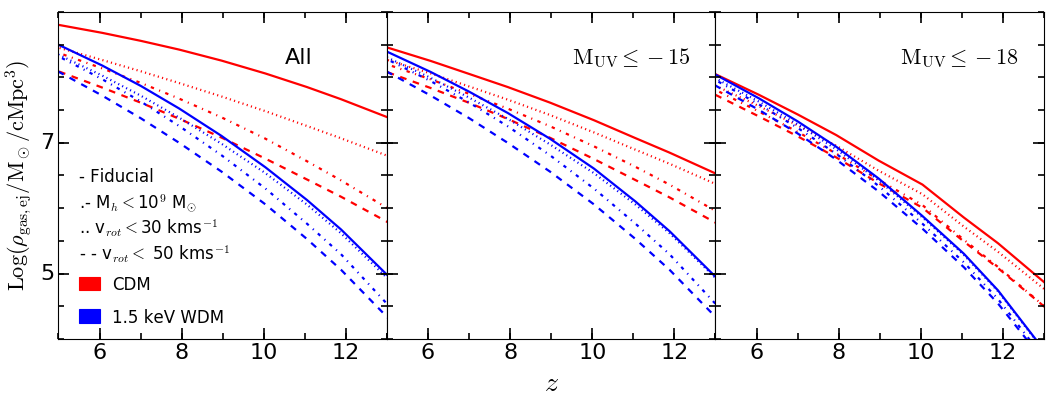}}
\caption{The ejected gas mass density as a function of redshift considering all galaxies (left panel), those with $\muv \lsim-15$ (middle panel) and $\muv\lsim-18$  (right panel). In each panel, the red and blue lines show results for CDM and 1.5 keV WDM, respectively, for the different feedback models noted in the legend.}
\label{fig_ejgasd} 
\end{figure*}

Starting with CDM, we find that the SMD smoothly grows with decreasing redshift as a larger number of galaxies assemble their stellar mass in a given volume. For the fiducial case, the SMD value grows by about two orders of magnitude ($10^{5.75-7.5}\Msun\, \mpc^{-3}$) over the 800 Myrs between $z \simeq 13$ and 5 as shown in Fig. \ref{fig_smd}. The SMD value decreases with the addition of UV feedback at all $z$ as the baryonic content of low mass galaxies is progressively suppressed; again, the impact successively increases from a cut-off of $v_{circ} = 30\, \kms$ to $M_h =10^9 \Msun$ to $v_{circ} = 50\, \kms$. With decreasing redshift larger systems assemble for which most of the stellar mass is built-up by a combination of in-situ star formation and mergers of progenitors above the UV suppression mass. This naturally results in a steeper $z$-evolution of the SMD with increasing UV feedback - indeed, compared to the fiducial case, galaxies in the ``maximal" UV feedback scenario with $v_{circ} = 50\, \kms$ assemble only about $11\%$ of the SMD at $z \simeq 13$, that rises to $\sim 66\%$ by $z \simeq 5$. Both the value of the SMD and the impact of UV feedback decrease when only considering galaxies brighter than a limit of $\muv =-15$ which provide roughly $30\%$ of the SMD at $z \simeq 13$ in the fiducial model rising to about $78\%$ by $z \simeq 5$. As expected, $\muv \lsim -18$ galaxies, that contribute $\sim 1\%\, (46\%)$ to the total SMD at $z \simeq 13\, (5)$ are impervious to the effects of UV feedback. 

The 1.5 keV WDM model shows a much steeper $z$-evolution of the SMD compared to CDM, irrespective of the feedback prescription used for the latter which is the result of two effects: an intrinsic dearth of low mass halos and a faster baryonic assembly since WDM galaxies start from larger progenitors that are less feedback limited \citep[see also][]{dayal2014}. Indeed comparing fiducial models, all the galaxies in the 1.5 keV WDM model contain less than 1\% of the total SMD at $z \simeq 13$ compared to CDM, thereafter rising steeply to the CDM value at $z \simeq 5$. As expected, the gap between CDM and 1.5 keV WDM SMDs decreases as we consider progressively massive systems with $\muv \lsim -15$ and as bright as $\muv \lsim -18$. It is interesting to note that, given its lack of low-mass halos, the 1.5 keV WDM model is much less affected by UV feedback - the difference between the fiducial and maximal UV feedback models is almost constant at $\lsim 0.3$ dex compared to the $\sim 1$ dex seen for CDM for $\muv \gsim -15$ galaxies. 

We reiterate the result found in  \citep{dayal2014} - that the $z$-evolution of the SMD is steeper in the 1.5 keV WDM model, irrespective of the baryonic feedback model considered. The $z$-evolution of the SMD, integrating down to magnitudes as faint as $-16.5$ with the JWST, can therefore be a powerful probe of the nature of DM. 

\subsection{Feedback impact on the ejected gas mass density}
\label{sec_gasejd}
Now that our model results, for both CDM and 1.5 keV WDM, have been shown to match existing observations, we study the impact of feedback on the total ejected gas mass density integrated over the entire history of all galaxies - $\rho_{gas,ej}$. Given our assumption of perfect metal-mixing in the ISM, $\rho_{gas,ej}$ is an excellent tracer of the metal enrichment of the IGM, as discussed in Sec. \ref{sec_IGMenrichment} that follows. 

\begin{figure*}
\center{\includegraphics[scale=0.36]{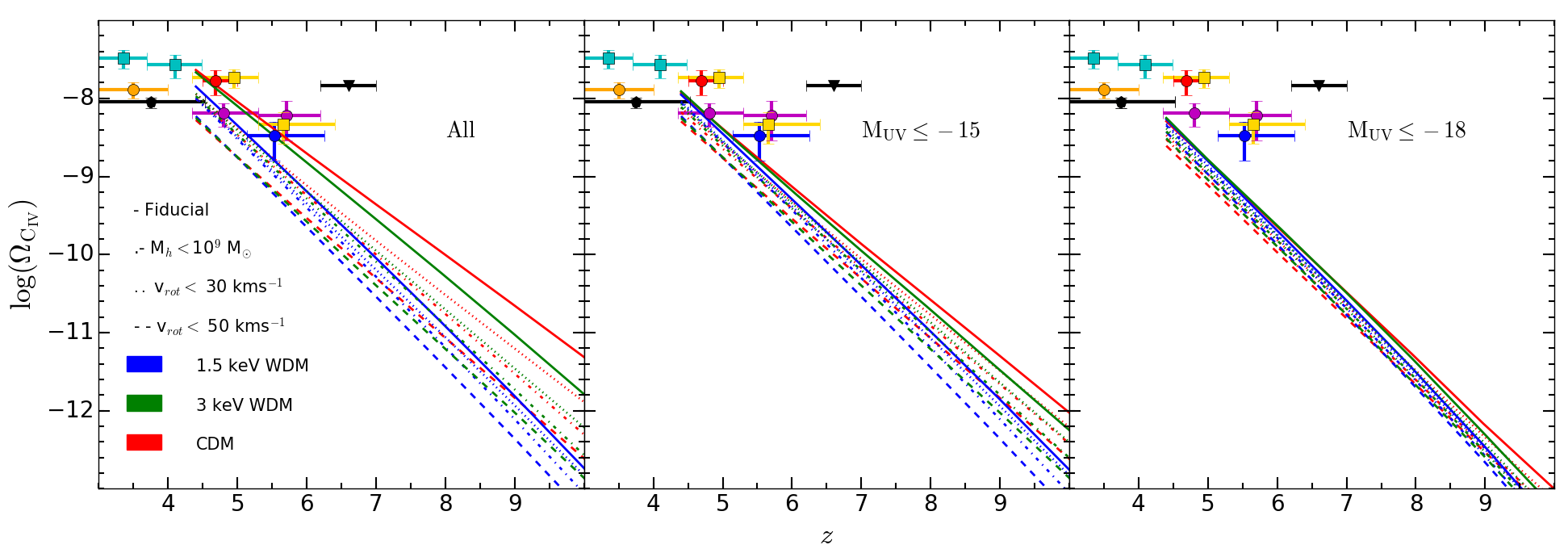}}
\caption{The cosmic mass density of CIV, $\Omega_{\mathrm{CIV}}$, measured as a function of redshift assuming all galaxies to have $Z_{gas}=0.2Z_{\odot}$, independent of mass and redshift, for all galaxies (left panel), galaxies with $\muv \lsim -15$ (middle panel) and $\muv \lsim-18$  (right panel). In each panel, the red, green and blue lines show results for CDM, 3 keV WDM and 1.5 keV WDM, respectively, for the different feedback models noted in the legend. Points indicate the CIV density parameter inferred observationally by: \citet[][rescaled by \citealt{ryanweber2009}, red circle]{pettini2003}, \citet[][orange circle]{dodorico2010}, \citet[][gold squares]{simcoe2011}, \citet[][black pentagon]{cooksey2013}, \citet[][magenta circles]{dodorico2013}, \citet[][cyan squares]{boksenberg2015}, \citet[][blue circle]{diaz2016} and \citet[][black triangle showing upper limit]{Bosman2017}.}
\label{CIV} 
\end{figure*}

Starting by considering all galaxies in CDM, we find $\rho_{gas,ej}$ in the fiducial case is about 44 (20) times higher than the SMD at $z \simeq 13\, (5)$ indicating the enormous impact of SNII feedback in ejecting gas from the potential wells of low-mass halos. As in the SMD studied above, the complete suppression of baryonic mass leads to a decrease in the 
ejected gas mass density when using a cut-off of $v_{circ} = 30\, \kms$ to $M_h =10^9 \Msun$ to $v_{circ} = 50\, \kms$. Using a UV feedback cut-off value of $M_h =10^9 \Msun (v_{circ} = 50\, \kms)$ results in $\rho_{gas,ej}$ decreasing by a factor of 40 (25) at $z \simeq 13$, reducing to a factor of $3$ (5) by $z \simeq 5$. As expected, the value of $\rho_{gas,ej}$ progressively decreases when considering galaxies with $\muv \lsim -15$ and $\muv \lsim -18$. Comparing values in the fiducial models, galaxies brighter than a magnitude limit of $\muv \lsim -15\, (-18)$ only contribute about $13\, (0.2)\%$ to the total $\rho_{gas,ej}$ value at $z \simeq 13$ that rises to about $45\, (18)\%$ by $z \simeq 5$, implying that the most ejected gas mass comes from galaxies fainter than $\muv =-15$ in CDM. Naturally, given the suppression of the baryonic component of low mass halos, including UV feedback results in a smaller difference when comparing $\rho_{gas,ej}$ from all galaxies to those above a certain magnitude cut. We also note that the difference between $\rho_{gas,ej}$ values for the fiducial and UV feedback models decreases when only considering relatively bright galaxies from about 1.6 dex for all galaxies to about $0.8 \, (0.4)$ dex for $\muv \lsim -15\, (-18)$ at $z \simeq 13$. 

As for the 1.5 keV WDM, a dearth of low mass halos leads to a lower $\rho_{gas,ej}$ value compared with CDM in any feedback scenario at $z \gsim 9$ with most ($\sim 79\%$) of the ejected gas mass density now being contributed by galaxies brighter than $\muv=-15$ at $z\approx 5$. Further, the $\rho_{gas,ej}$ trend flips at lower-$z$ with 1.5 keV WDM models that include UV feedback having a larger ejected gas mass density value compared to the corresponding CDM model. Analogous to the steeper build-up of the SMD discussed above, this is a result of galaxies starting from larger, and hence less feedback-limited, progenitors in 1.5 keV WDM that have higher star formation rates leading to a larger ejection of gas mass at later epochs. As also noted for the SMD, we see that the difference between the fiducial and UV feedback limited $\rho_{gas,ej}$ values is roughly constant at $\sim 0.5$ dex, compared to the larger and $z$-dependent values seen for CDM, with the differences being of the order of 0.2 dex for a magnitude cut of $\muv \lsim -18$. Finally, we note that the relative CDM and 1.5 keV trends discussed here imply a delayed but accelerated IGM metal-enrichment scenario in the latter model as studied in Sec. \ref{sec_IGMenrichment} that follows.

\section{The IGM metal-enrichment in CDM and WDM and comparison with observations}
\label{sec_IGMenrichment}

We now use the ejected gas mass density values, calculated above, to obtain an estimate of the IGM metal enrichment in the two metallicity scenarios adopted in this work: the first where the gas-phase metallicity $Z_{gas}=0.2\Zsun$ for all galaxies and the second where $Z_{gas}$ for a given galaxy is computed depending on its stellar mass. Given that the CIV content, estimated from quasar absorption lines, is used as an indicator of the IGM metal enrichment ($\Omega_{\mathrm{CIV}}$), we convert our values of the gas mass density ejected by a galaxy into the CIV density parameter using $\Omega_{\mathrm{CIV}} = \rho_{\mathrm{CIV}}/\rho_c$. Here $\rho_{\mathrm{CIV}}$ and $\rho_c$ represent the CIV and critical densities, respectively. Further, $\rho_{\mathrm{CIV}}$ is calculated by summing over the gas mass ejected by all the, say $N$, galaxies at a given $z$ such that 
\begin{equation}
\rho_{\mathrm{CIV}} = \sum_{i=1}^{N} \rho_{gas,ej}(i) \times Z_{gas}(i) \times f(\mathrm{C/Z}) \times f(\mathrm{CIV/C}),
\label{OMCIV}
\end{equation}
where for each galaxy ($i$) $\rho_{gas,ej}(i)$ is the total gas mass density ejected by the galaxy over its lifetime till $z$ and $Z_{gas}$ is the metallicity of the perfectly-mixed ISM gas. Further, $f(\mathrm{C/Z})$ is the fraction of metals in the form of carbon and $f(\mathrm{CIV/C})$ represents the fraction of triply ionized Carbon. Assuming SNII to be the main dust sources, the value of $f(\mathrm{C/Z})$ is obtained by extrapolating the SNII yields (between $13-40\Msun$) given by \citet{nomoto2006} down to $8 \Msun$ and weighting these over a Salpeter IMF between $8-40\Msun$; stars with mass $\gsim 40\Msun$ collapse to Black Holes with little contribution to the metal budget. This calculation results in a value of $f(\mathrm{C/Z}) \simeq 0.14$. We use the results from \citet{keating2016} and \citet{garcia2017b} to find ${\rm Log} (\mathrm{CIV/C}) = -0.35(z+1)+1.45$ for $z \gsim 4$, yielding $f(\mathrm{CIV/C}) \simeq 0.5$ at $z=4$, consistent with observations and photometric modeling by \citet{simcoe2011b}, that decreases to $f(\mathrm{CIV/C}) \simeq 0.009$ by $z=9$. We note that in using the $f(\mathrm{C/Z})$ yield purely from SNII, we have neglected the metal contribution from metal-free (Pop III) stars. This is justified by the fact that observations of high-$z$ UV slopes \citep{dunlop2013, rogers2013,rogers2014,bouwens2013,oesch2016} and star formation clumps \citep{vanzella2017} show no indication of metal-free stellar populations, a result that is supported by theoretical simulations that find PopIII stars to contribute $\leq 10 \%$ to star formation at $z \leq 7-10$ \citep{tornatore20072,maio2010,pallotini2014,jaacks2018} and $< 5 \%$ to the luminosity for galaxies with M$_{\mathrm{UV}} <-16$ at $z=10$ \citep{salvaterra2011}. Furthermore, the observed ratios of CII, OI, SiII and FeII in quasar absorption line systems at $4.7<z<6.3$ show no differences with respect to metal-poor systems at lower redshifts \citep{becker2012}.

We start with the simplest scenario where each galaxy has a fixed metallicity of $Z_{gas}=0.2 \Zsun$. This assumption likely over-estimates (under-estimates) the metallicity values for low-mass galaxies at high-$z$ (high-mass galaxies at low-$z$).  The $\Omega_{\mathrm{CIV}}(z)$ values arising from these calculations for CDM, 1.5 and 3 keV WDM are shown in Fig. \ref{CIV}.

\begin{figure*}
\center{\includegraphics[scale=0.34]{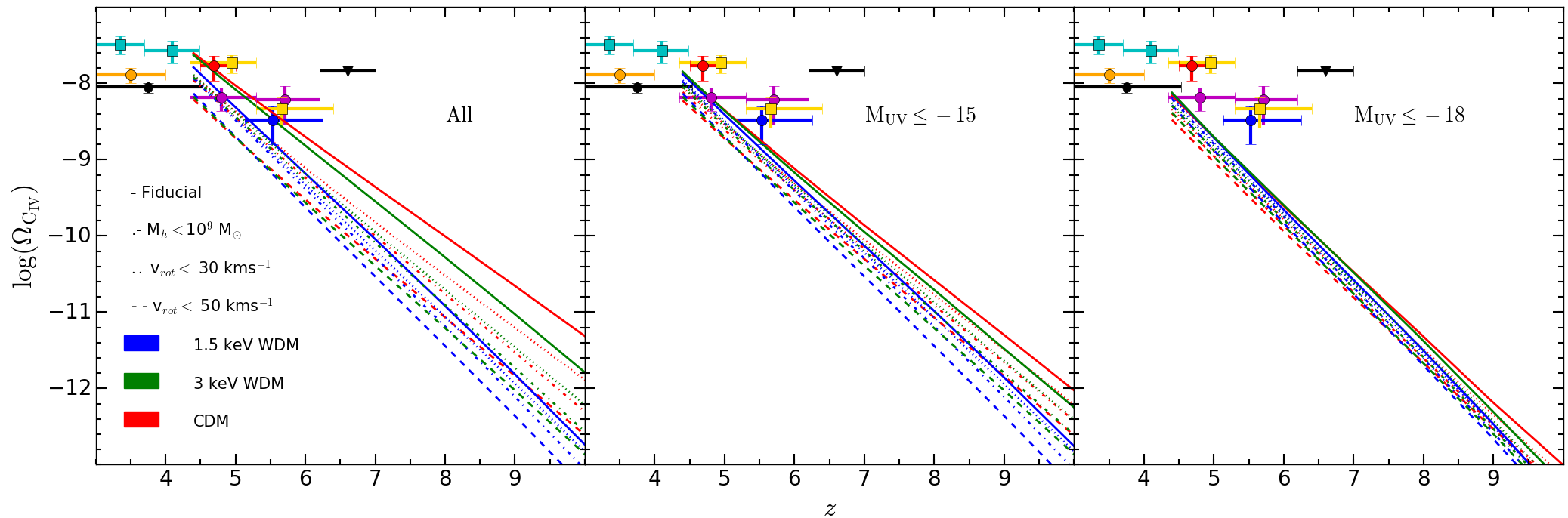}}
\caption{The cosmic mass density of CIV, $\Omega_{\mathrm{CIV}}$, measured as a function of redshift assuming all galaxies to have $Z_{gas}=fn(M_{*})$. Results are shown for all galaxies (left panel), galaxies with $\muv \lsim -15$ (middle panel) and $\muv \lsim-18$  (right panel). In each panel, the red and blue lines show results for CDM and 1.5 keV WDM, respectively, for the different feedback models noted in the legend. Points indicate observational data for which the references are shown in the caption of Fig. \ref{CIV}.
}
\label{CIVvariations} 
\end{figure*}

We focus on comparing our results, for CDM, 3 and 1.5 keV WDM, to the $\Omega_{\mathrm{CIV}}$ observational data at $z \simeq 5.5$ given that metal enrichment from Asymptotic Giant Branch (AGB) stars, which we have neglected in our calculations, could have had a significant contribution at lower $z$; we note that we have used the same baryonic free parameter values for all three models. We find that the CDM and 3 keV WDM fiducial models where all galaxies contribute to the IGM metal enrichment agree with the observational data points of \citet{simcoe2011} and \citet[][that supersedes \citealt{ryanweber2009}]{diaz2016}. Within error bars, the \citet{diaz2016} point, with the lowest measured $\Omega_{\mathrm{CIV}}$ value at $z \sim 5.5$, also matches the CDM model with complete UV suppression in 
galaxies with $v_{circ} < 30\, \kms$ as well as the fiducial 1.5 keV WDM model. The intermediate \citet{simcoe2011} point rules out all models except fiducial CDM and 3 keV at $\gsim 1.6-\sigma$. On the other hand, with its highest measured value of $\Omega_{\mathrm{CIV}}$ at $z \sim 5.5$, the \citet{dodorico2013} point only allows the CDM fiducial model, ruling out the 3 keV WDM fiducial model (all other models) at $\approx 1.1\sigma$ ($\gsim 1.5-\sigma$).

As for the key metal polluters, our results show that, in the fiducial model, galaxies with $\muv\gsim -15$ ($\muv \lsim -15$) could provide roughly 50\% (80\%) of the IGM metal budget in CDM (1.5 keV WDM) model by $z \simeq 4.5$. As expected, the currently detected brighter galaxies, with $\muv\lsim -18$, have a smaller contribution of about 22\% (38\%) to the metal budget for CDM (1.5 keV WDM); the results from the 3 keV model naturally lie between CDM and 1.5 keV WDM. 

Parameterizing the $\Omega_{\mathrm{CIV}}-z$ relation as log($\Omega_{\mathrm{CIV}})=a(1+z)+b$, we show the slopes for all CDM and 1.5 keV WDM models in Table \ref{OMCIVpars}. We start by noting that the steeper $z$-evolution of $\rho_{gas,ej}$ in 1.5 keV WDM with respect to CDM is reflected in its steeper (by a factor of $1.3$) $\Omega_{\mathrm{CIV}}-z$ relation - the fiducial CDM model predicts a 27 times higher value of CIV compared to the fiducial 1.5 keV model at $z \simeq 10$, reducing to a factor of about $2$ by $z=5$. Given the lack of low-mass halos, the impact of UV feedback is naturally lesser on the 1.5 keV WDM model as compared to CDM, resulting in a larger steepening of CDM slopes. As shown in the same table, the CDM slopes are shallower by a factor of $a \sim 1.1-1.2$ when compared to 1.5 keV WDM. 

While, as expected, the CDM fiducial model shows the highest value of $\Omega_{\mathrm{CIV}}$, these results show a degeneracy between the underlying DM model and the baryonic feedback prescription implemented. This highlights the fact that an intrinsic dearth of low mass halos (in light WDM models) is equivalent to increasing the UV feedback thereby suppressing, the baryonic content and star formation capabilities of, low-mass halos in CDM. For example, at $z \simeq 5.5-9$, the 1.5 keV WDM fiducial model lies between the CDM models with UV suppression limits of $v_{circ} \lsim 30\, \kms$ and $v_{circ} \lsim 50\, \kms$, analogous to the UV LF behaviour seen in Sec. \ref{sec_uvlf}. 

In order to check the dependence of our results on the assumed metallicity, we explore an alternative scenario in which the gas-phase metallicity scales with the stellar mass. This assumption is motivated by the observed mass-metallicity relation (MZR) linking the gas-phase metallicity and stellar mass from $z =0$ to $\sim 4$ \citep[][]{tremonti2004,lee2006,maiolino2008,mannucci2009,zahid2012,hunt2016}. 
For this work, we use the results, at the highest measured redshifts of $z=3-4$, from the LSD and AMAZE surveys  \citep{maiolino2008,mannucci2009} which can be fit to yield 
${\rm log}(Z_{gas}/Z_{\odot}) = 0.383 \, {\rm log}(M_{*})-4.307$ for galaxies with $M_{*}\gsim 10^{9.4}M_{\odot}$; we assume each galaxy to have $Z_{gas} = 0.20\, Z_{\odot}$ below this mass range \footnote{Using a lower value of $Z_{gas} = 0.10\, Z_{\odot}$ results in all models under-predicting the $\Omega_{\mathrm{CIV}}$ values as compared to observations at $z \gsim 4.5$. However, this result in not unreasonable given our assumption of metals being homogeneously distributed in the IGM.}. We use Eqn. \ref{OMCIV} to recompute the value of $\Omega_{\mathrm{CIV}}(z)$ using this $M_*$-dependent metallicity, the results of which are shown in Fig. \ref{CIVvariations} and in Table \ref{OMCIVpars}. Interestingly, we find these results to be indistinguishable, in terms of the $\Omega_{\mathrm{CIV}}$ values, from those assuming a constant metallicity of $Z_{gas} = 0.20\, Z_{\odot}$: this is driven by the fact that 
low-mass galaxies, which are the key contributors to the ejected gas mass density as shown in Sec. \ref{sec_gasejd}, are assumed to have the same gas-phase metallicity in both the models considered here. However, given the larger metallicities of massive galaxies in this latter calculation results in massive galaxies ($\muv\lsim -18$) having a larger contribution the IGM metal budget: in the fiducial CDM (1.5 keV WDM) model, these galaxies contribute $28\%\ (46\%)$ to the IGM metal budget by $z \simeq 4.5$ as compared to the slightly lower values of $22\%\ (38\%)$ assuming a constant metallicity of $0.2\Zsun$. Critically, we find that assuming a $M_*$-dependent metallicity has no sensible impact on the $\Omega_{\mathrm{CIV}}-z$ relation for any of the CDM or 1.5 keV WDM models or their relative differences, both including/excluding the impact of UV feedback.
   
We note that our calculations have involved a number of simplifications which are now summarized: (i) all metals are assumed to be perfectly mixed with gas implying outflows to have the same metallicity as the ISM gas; (ii) at any $z$ we assume at least the lowest mass galaxies ($M_{*}\lsim 10^{9.4}M_{\odot}$) to have a fixed gas metallicity of $Z_{gas}=0.2 \Zsun$ which is, most likely, an over-estimation at the highest redshifts; (iii) we use a halo mass independent $\c4/C$ ratio to which the CIV density is sensitive; (iv) we have only considered Carbon yields from SNII, neglecting the contribution from AGB stars that would have a significant impact, specially at $z \lsim 5$ at which the metal mass would be underestimated; (v) while metals should be concentrated in over-dense regions, we assume them to be homogeneously distributed over the IGM in order to infer the $\Omega_{\c4}$ value; and (vi) $Z_{gas}$, and in turn the extent to which the IGM is polluted with metals critically depends on the metallicity of inflowing and outflowing gas: outflows preferentially carrying away metals can lead to an enhanced IGM metallicity enrichment whilst lowering the ISM metallicity. On the other hand, inflows of metal-poor gas can dilute the ISM metallicity whilst inflows of metal-enriched gas, possibly previously ejected by the galaxy (the so-called ``galactic fountain") can increase the ISM metallicity. Whilst assuming perfect mixing in this case results in a lower (higher) IGM metallicity in these two scenarios, respectively, relaxing this assumption can either enhance/decrease the IGM metallicity, depending on the metal-richness (metal to gas ratio) of the outflows. However, accounting for such non-linear effects requires simultaneously, and consistently, modelling the metal-cycle in the ISM and IGM which, extending much beyond the scope of this proof-of-concept paper, is deferred to future works.

 
At this point, in addition to the metal-cycle and baryon prescription-cosmology degeneracies discussed above, we highlight other key degeneracies that could lead to similar physical scenarios: firstly, the metallicity of outflowing gas has a degeneracy with the fractional volume of the IGM polluted with metals i.e. a given value of the IGM metallicity can be obtained by polluting a small (large) fraction of the IGM with low (high) metallicity gas. However, this calculation is extremely hard to carry out without modelling both the metal enrichment and metal dispersion calculations in the IGM. Furthermore, it must be noted that the ``average" value of the IGM metallicity is hard to obtain observationally given it is only measured along a few lines of sight. A second degeneracy that can arise in such calculations is cosmology dependent: given that CDM collapses on all scales, the clumping factor (over-density above average) of the IGM is expected to be higher than for WDM where low-$\sigma$ density fluctuations can get wiped out. Reasonably assuming metal pollution to percolate more easily in low-density regions, this implies that the IGM in CDM could have a lesser volume (of denser gas) metal enriched to a higher level than WDM assuming the same amount of metals ejected into the IGM. However, this patchy metal enrichment could possibly be countered by the more homogeneous galaxy distribution in CDM as opposed to the larger galaxy bias expected in WDM. However, such calculations require, both, spatial information of galaxy positions as well as jointly tracking the baryonic assembly and metal exchange between the ISM and IGM that we defer to future works.

\begin{table*}
\centering
\caption {Parameterizing the $\Omega_{\mathrm{CIV}}$-$z$ relation as log$(\Omega_{\mathrm{CIV}}) = a(1+z)+b$, we show the slopes ($a$) for all CDM and 1.5 keV WDM models for the two cases considered in Sec. \ref{sec_IGMenrichment}: the first where $Z_{gas}=0.20Z_{\odot}$ and the second where $Z_{gas}=fn(M_{*})$.} 
\label{OMCIVpars}
\centerlast
\begin{tabular}{ |c|c|c|c|c|}
 \hline
  DM model &  Fiducial model & $\mathrm{M_{h}<10^{9}M_{\odot}}$ & $\mathrm{v_{c}<30kms^{-1}}$  & $\mathrm{v_{c}<50km^{-1}}$  \\
 \hline
 \multicolumn{5}{|c|}{Slopes ($a$) for $Z_{gas}=0.20Z_{\odot}$  }\\
 \hline
CDM  & -0.66  &-0.77  &-0.70&-0.77 \\
1.5 keV WDM & -0.87 & -0.90 & -0.86 &-0.90 \\
 \hline
  \multicolumn{5}{|c|}{Slopes ($a$) for $Z_{gas}=fn(M_{*})$ }\\
 \hline
 CDM  & -0.66  & -0.77 &-0.71 & -0.78\\
1.5 keV WDM & -0.88 &-0.91 & -0.87 &-0.92 \\
 \hline
 \end{tabular}
\end{table*} 

\section{Conclusions and discussion}
This {\it proof-of-concept} work focuses on studying the metal enrichment of the IGM in cold and warm dark matter (1.5 keV) cosmologies using {\it Delphi} - a semi-analytic model \citep{dayal2014, dayal2014_wdm1, dayal2017, dayal2017b} that jointly tracks the DM and baryonic assembly of high-redshift ($z \gsim 4$) galaxies. This work is motivated by the fact that, compared to CDM, 1.5 keV WDM has a significant fraction ($\gsim 95\%$) of bound DM mass missing in low mass halos ($M_h \lsim10^{9.5}\mathrm{M_{\odot}}$) at any cosmic epoch - this loss of shallow potential wells, expected to be the key IGM metal-polluters, would naturally result in a delayed and lower metal enrichment in 1.5 keV WDM when compared to CDM. In addition to the {\it fiducial} (SNII feedback only) model, we explore three ``maximal" scenarios for reionization feedback by completely suppressing the gas mass, and hence star formation capabilities, in all halos below (i) $M_h = 10^9 \Msun$; (ii) $v_{circ} = 30\, \kms$; and (iii) $v_{circ} = 50\, \kms$. The model uses two mass- and $z$-independent free parameters - the fraction of SNII energy coupling to the gas ($f_w$) and the instantaneous star formation efficiency ($f_{*}$) to capture the key physics driving early galaxies. These are calibrated to the observed UV LF at $z \simeq 5-10$ yielding $f_w=$ 10\% and $f_* = 3.5\%$ for the fiducial model and we use the same parameter values for all models.

We find that while the latest LBG UV LFs \citep{bouwens2016b, livermore2017} are consistent with CDM and the 3 keV and 1.5 keV fiducial (SNII feedback only) models, they allow ruling out maximal UV feedback suppression below $v_{circ} = 50\, \kms$ for CDM and {\it all maximal} UV feedback models for 1.5 keV WDM. However, given that it is only measured for massive $\muv \lsim -18$ galaxies, as of now, all models are compatible with the SMD - as noted in previous works, the SMD will have to be measured down to magnitudes as faint as $\muv=-16.5$, with e.g. the JWST, to be able to distinguish between CDM and 1.5 keV WDM \citep[e.g.][]{dayal2014}. In terms of the total ejected gas mass density, we find that while galaxies fainter than $\muv = -15$ contribute most ($\sim 55\%$) to this quantity in CDM at $z=5$, the trend reverses with $\muv \lsim -15$ galaxies dominating in 1.5 keV WDM.

We explore two gas-phase metallicity scenarios: one where all galaxies have a constant gas-phase metallicity of $Z_{gas}=0.2Z_{\odot}$ and the other in which we assign metallicities using the $z \sim 3-4$ MZR for galaxies with $M_* \gsim 10^{9.4}\Msun$ with lower mass galaxies assumed to have a fixed metallicity of $Z_{gas}=0.2Z_{\odot}$. Assuming all galaxies to have a constant gas-phase metallicity of $Z_{gas}=0.2Z_{\odot}$, a natural consequence is that $\muv \gsim -15$ ($\muv \lsim -15$) galaxies are the key IGM metal polluters in CDM (1.5 keV WDM), contributing $\sim 50\%$ ($80\%$) to the total IGM metal budget at $z \simeq 4.5$ with currently detected galaxies ($\muv\lsim -18$) contributing $\sim 22\%\,  (38\%)$ to the IGM metal budget; applying the mass-metallicity relation observed at the highest redshifts of $z \sim 3-4$ yields qualitatively similar results, with the metal contribution from observed galaxies increasing slightly to $28\%\ (46\%)$ in the fiducial CDM (1.5 keV WDM) model.  

Independent of the two gas-phase metallicity models assumed in this work, current observations on the IGM metal budget, obtained through measurements of $\Omega_{\c4}$, specially at $z \sim 5.5$, allow the following constraints: while, within its $1-\sigma$ error bars, the \citet{diaz2016} point is consistent with both the fiducial and maximal reionization feedback (suppressing all halos below $v_{circ} = 30\, \kms$) models for CDM and the 3 and 1.5 keV WDM fiducial models, the \citet{simcoe2011} point rules out {\it all models except fiducial CDM and 3 keV at $> 1.6-\sigma$}. Our results therefore imply that, combining the two different data sets provided by the evolving UV LF and IGM metal density \citep{simcoe2011, dodorico2013}, we can effectively rule out all models other than fiducial CDM; a combination of the UV LF and the \citet{diaz2016} points provides a weaker constraint, allowing fiducial CDM and the 3 and 1.5 keV WDM models, as well as CDM with UV suppression of all halos with $v_{circ} \lsim 30\, \kms$. Tightening the error bars on $\Omega_{\c4}$, future observations at $z \gsim 5.5$ could therefore well allow ruling out WDM as light as 1.5 keV.

\section*{Acknowledgments} 
JB \& PD acknowledge support from the European Research Council's starting grant ERC StG-717001 ``DELPHI". PD acknowledges support from the European Commission's and University of Groningen's CO-FUND Rosalind Franklin program. ERW acknowledges the support of Australian Research Council grant DP1095600.


\bibliographystyle{mn2e}
\bibliography{wdm_v2}

\label{lastpage} 
\end{document}